\documentclass[reprint, double column, superscriptaddress,prl, showkeys]{revtex4-1}

\usepackage{graphicx,epsfig}
\usepackage{amssymb}
\usepackage{amsmath}
\usepackage{bm}
\usepackage{textcomp}
\usepackage{color}
\usepackage{pgffor}

\usepackage{soul}

\usepackage{pdfpages}
\usepackage{BOONDOX-cal}

\makeatletter

\AtBeginDocument{\let\LS@rot\@undefined}
\makeatother

\newif\ifarXiv
\arXivtrue

\begin{document}
\setcounter{page}{1}

\title[]{Robust quantum Hall ferromagnetism near a gate-tuned $\nu=1$ Landau level crossing}
\author{Meng K. \surname{Ma}}
\thanks{These two authors contributed equally}
\author{Chengyu \surname{Wang}}
\thanks{These two authors contributed equally}
\author{Y. J. \surname{Chung}}
\author{L. N. \surname{Pfeiffer}}
\author{K. W. \surname{West}}
\author{K. W. \surname{Baldwin}}
\affiliation{Department of Electrical and Computer Engineering, Princeton University, Princeton, New Jersey 08544, USA}
\author{R. \surname{Winkler}}
\affiliation{Department of Physics, Northern Illinois University, DeKalb, Illinois 60115, USA}
\author{M. \surname{Shayegan}}
\affiliation{Department of Electrical and Computer Engineering, Princeton University, Princeton, New Jersey 08544, USA}
\date{\today}

\begin{abstract}

In a low-disorder two-dimensional electron system, when two Landau levels of opposite spin or pseudo-spin cross at the Fermi level, the dominance of the exchange energy can lead to a ferromagnetic, quantum Hall ground state whose gap is determined by the exchange energy and has skyrmions as its excitations. This is normally achieved via applying either hydrostatic pressure or uniaxial strain. We study here a very high-quality, low-density, two-dimensional hole system, confined to a $30$-nm-wide (001) GaAs quantum well, in which the two lowest-energy Landau levels can be gate tuned to cross at and near filling factor $\nu=1$. As we tune the field position of the crossing from one side of $\nu=1$ to the other by changing the hole density, the energy gap for the quantum Hall state at $\nu=1$ remains exceptionally large, and only shows a small dip near the crossing. The gap overall follows a $\sqrt{B}$ dependence, expected for the exchange energy. Our data are consistent with a robust quantum Hall ferromagnet as the ground state.

\end{abstract}

\maketitle  

Quantum ferromagnetism of itinerant electrons has been a subject of interest for decades \cite{Bloch.ZP.1929, Stoner.RPP.1947, Ashcroft.Mermin.1975}. In its broadest context, when the electrons’ exchange energy dominates over the Fermi (kinetic) and disorder energies, the electrons should align their spins and form a ferromagnetic ground state. However, this criterion is extremely difficult to achieve in metallic elements because of the very large Fermi energies \cite{Ashcroft.Mermin.1975}. In flat-band and dilute two-dimensional (2D) carrier systems, the criterion has been met very recently and ferromagnetic ground states have been reported \cite{Sharpe.Science.2019, Roch.PRL.2020, Polshyn.Nature.2020, Hossain.PNAS.2020}, although it is important to keep in mind that disorder can play a very important role \cite{ Kim&Kivelson.PNAS.2021}. A ferromagnetic ground state has also been reported in a system of exotic, electron-magnetic-flux quasi-particles, namely composite fermions \cite{Hossain.Nat.Phys.2021}, and a spontaneous valley polarization of itinerant electrons has been observed in a dilute 2D electron system where electrons normally occupy two, degenerate conduction-band valleys \cite{Hossain.PRL.2021}; in the latter case, the electrons' valley degree of freedom plays the role of the spin degree of freedom.

Another platform for the emergence of quantum ferromagnetism is created when the 2D electron system is cooled to very low temperatures, and placed in a large perpendicular applied magnetic field ($B$). In this case, the electrons' kinetic energy is quenched as the electrons occupy the quantized Landau levels (LLs). In the simplest scenario, the main energy competing with the exchange energy is disorder and, in a sufficiently clean (low-disorder) electron system, the so-called quantum Hall (QH) ferromagnetism prevails. In particular, when two LLs of opposite spin polarity, or more generally pseudo-spin polarity, cross at the Fermi level, the 2D carriers form a polarized, QH ground state with an energy gap separating it from its excitations \cite{Ando.JPSJ.1974, Giuliani.PRB.1985, Sondhi.PRB.1993, Lay.APL.1993, Koch.PRB.1993, Maude.PRL.1996, Girvin&MacDonald.Perspectives.1996, Deneshvar.PRL.1997, Jungwirth.PRL.1998, Piazza.Nature.1999, DePoortere.Science.2000, Girvin.Phys.Today.2000, Jungwirth.PRB.2000, Muraki.PRL.2001, Shkolnikov.PRL.2002, Kellog.PRL.2004, Tutuc.PRL.2004, Eisenstein.Nature.2004, Shkolnikov.PRL.2005, Zhang.PRL.2005, Lai.PRL.2006, Vakili.PRL.2006, Zhang.PRL.2006, Nomura.PRL.2006, Padmanabhan.PRL.2010, Gokmen.PRB.2010, Young.Nat.Phys.2012, Parameswaran.JPCM.2019, Li.PRL.2020, Lupatini.PRL.2020}. The energy gap of such a QH ferromagnet is directly related to the Coulomb exchange energy, and the lowest-energy excitations can involve skyrmions, smooth spin textures, if the Zeeman energy is sufficiently small \cite{Sondhi.PRB.1993, Maude.PRL.1996, Shkolnikov.PRL.2005, Gokmen.PRB.2010, Barrett.PRL.1995}. The pseudo-spin degree of freedom includes the real spin, conduction-band valley, layer, or electric subband, depending on the nature of the crossing LLs. 

\begin{figure}[b!]
  \begin{center}
    \psfig{file=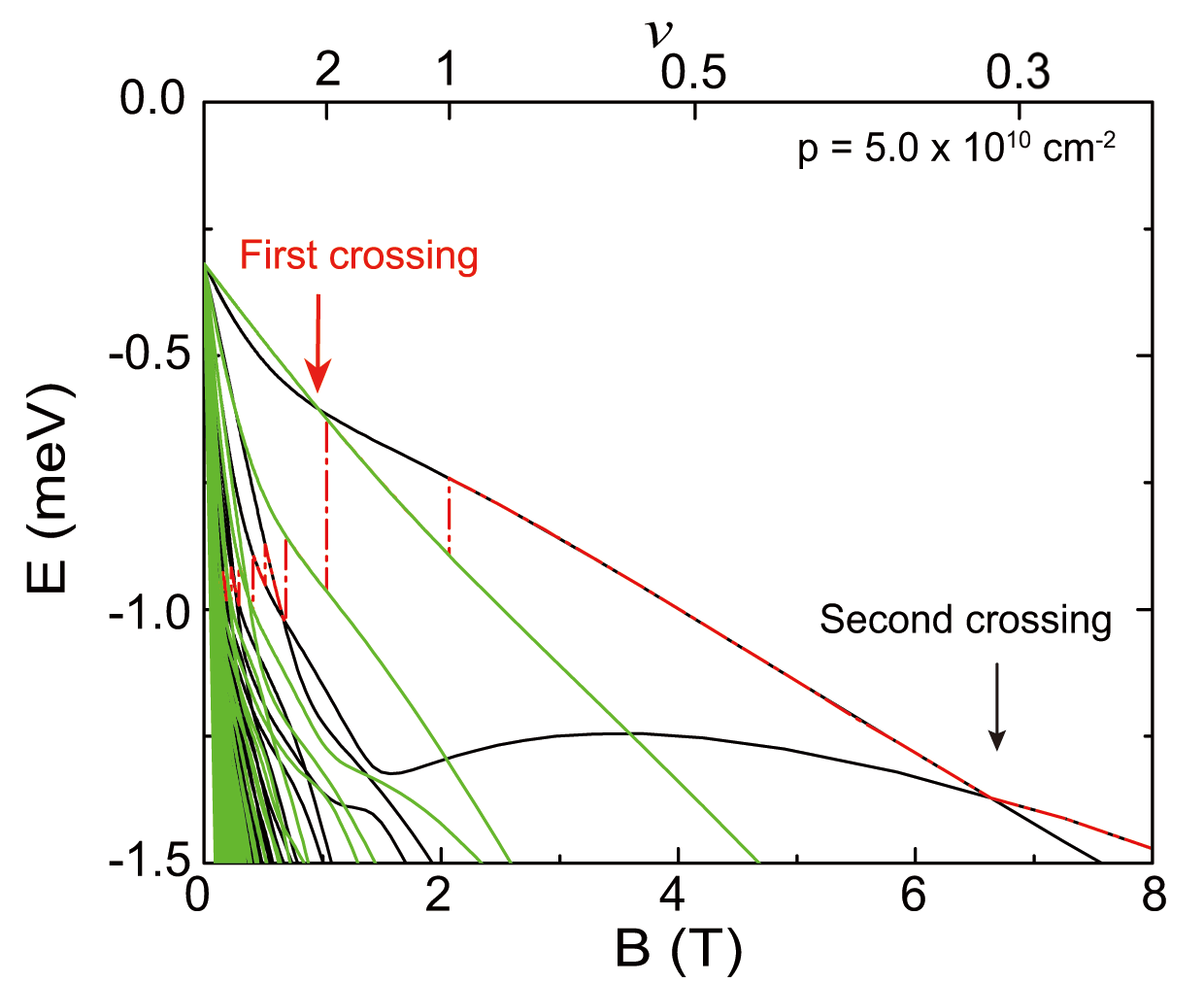, width=0.48\textwidth}
  \end{center}
  \caption{\label{LLcal}
   Calculated energy ($E$) vs. magnetic field ($B$) LL diagram for a 2DHS confined in a $30$-nm-wide, (001) GaAs QW at a density of $5.0\times10^{10}$ cm$^{-2}$. The upper axis indicates the filling factor $\nu$ which is proportional to $1/B$. Black and green lines indicate LLs of opposite effective spin \cite{Footnote.color.coding}, and the red dash-dotted line traces the Fermi energy. The vertical arrows mark the positions of the two crossings of the two lowest-energy (in magnitude) LLs.}
  \label{fig:LLCalc}
\end{figure}

\begin{figure*}
  \begin{center}
    \psfig{file=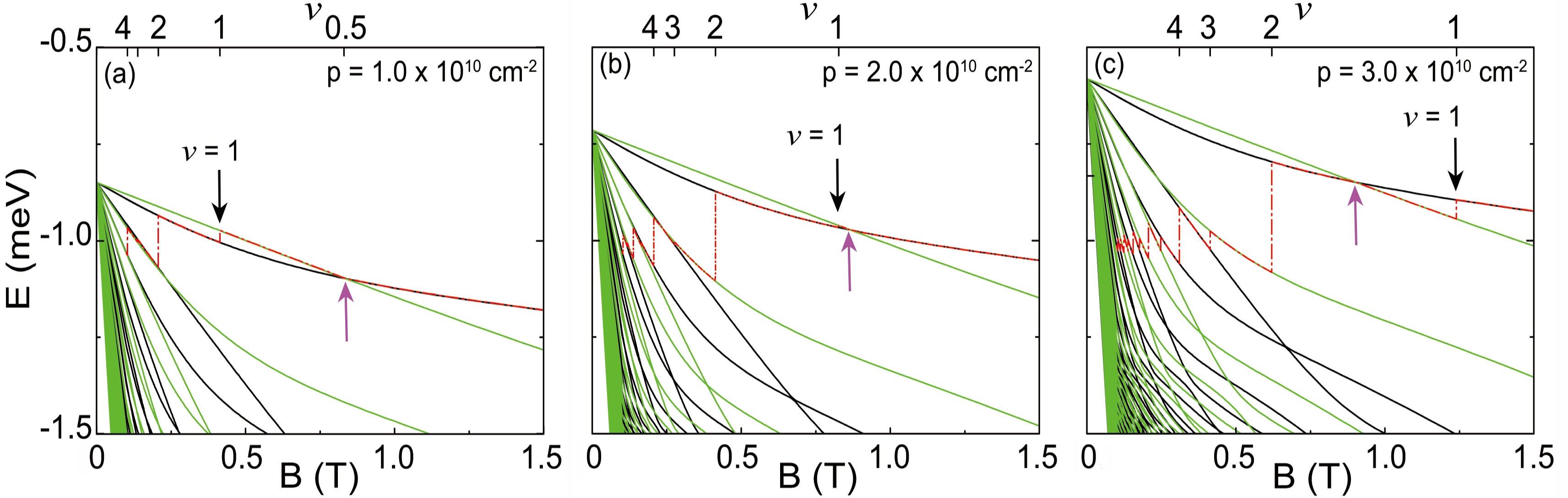, width=1\textwidth}
  \end{center}
  \caption{\label{LLcals} 
     Calculated LL diagrams for a 2DHS confined in a $30$-nm-wide (001) GaAs QW at densities: (a) $1.0$, (b) $2.0$ and (c) $3.0\times10^{10}$ cm$^{-2}$. The upper axes indicate the filling factor $\nu$. As in Fig. \ref{fig:LLCalc}, black and green lines indicate LLs of opposite effective spin \cite{Footnote.color.coding}, and the red dash-dotted line traces the Fermi energy. The downward arrows mark the positions of $\nu=1$, and the upward arrows the positions of the first crossing of the two lowest-energy LLs.
}
  \label{fig:LLcals}
\end{figure*}

Here we focus on the ground state of a very-high-quality, dilute, 2D \textit{hole} system (2DHS) confined to a (001) GaAs quantum well (QW) near LL filling factor $\nu=1$ as the two lowest-energy LLs cross (Fig. \ref{fig:LLCalc}). (In this paper, we implicitly refer to LLs with smaller \textit{magnitude} of energy, measured from the band edge, as having lower energy; in Fig. \ref{fig:LLCalc} and other figures, these LLs have the least negative energy values at any given $B$.)  The crossing of the two lowest-energy LLs is unusual and challenging to access experimentally. It can be achieved in 2D electron systems in GaAs via tuning the $g$-factor through zero by applying hydrostatic pressure \cite{Maude.PRL.1996}, or in AlAs via tuning the valley splitting energy by applying uniaxial strain \cite{Shkolnikov.PRL.2005}. The 2DHSs in GaAs provide a particularly rich platform in this context because, thanks to the strong spin-orbit interaction, their LLs are highly non-linear as a function of $B$ \cite{Winkler.Book.2003} and can exhibit multiple crossings, even at $\nu=1$. Moreover, the field positions of the crossings can be controlled by varying the QW width \cite{SM}, or simply by changing the 2D hole density.
 
To place our study in a broader perspective, in Fig. \ref{fig:LLCalc} we show the results of our self-consistently calculated 2D hole energy ($E$) vs. $B$ LLs at $p=5$, in units of $10^{10}$ cm$^{-2}$, which we use throughout this paper. The calculations were performed using the multiband envelope function approximation based on the $8 \times 8$ Kane Hamiltonian \cite{Winkler.Book.2003}. They are essentially a Hartree calculation and do not include the exchange interaction. Since pure spin is not a good quantum number in a system with strong spin-orbit interaction, we have grouped the LLs in Fig. \ref{fig:LLCalc} into two pseudo-spin species, color-coded using black and green \cite{Footnote.color.coding}. The red dash-dotted line traces the position of the Fermi energy ($E_F$) as a function of $B$. The lowest two LLs have the very unusual characteristic that, depending on the parameters of the 2DHS, they can cross \textit{twice} (Fig. \ref{fig:LLCalc}). The first crossing occurs between two LLs that have opposite pseudo-spin characters, while the second crossing involves a LL which emanates from an upper (excited) subband. The second crossing is remarkable on its own and entails very unusual evolutions of the ground state both at $\nu=1$ \cite{Graninger.PRL.2011, Liu.PRB.2015} as well as at $\nu=1/2$ where an even-denominator fractional QH liquid state \cite{Liu.PRB.2014}, or an anisotropic, pinned Wigner solid state \cite{Liu.PRL.2016} can emerge. The details of the first crossing, which is the subject of our investigation here, are less known \cite{Fischer.PRB.2007}. This is partly because the crossing typically occurs at small magnetic fields, implying that very dilute and yet low-disorder samples are needed for its study. Only very recently the presence of this crossing in a (001) GaAs 2DHS was demonstrated experimentally and, using optical techniques, it was shown that the 2DHS spins undergo a reversal as the LLs cross \cite{Lupatini.PRL.2020}.

Figure \ref{fig:LLcals} highlights in detail the expected evolution of the LLs and their crossings as a function of density. The calculations are shown for $p=1.0$, 2.0 and 3.0, and the color coding of the LLs is the same as in Fig. \ref{fig:LLCalc}. In all three panels, the two LLs with opposite pseudo-spin cross at $B$ slightly below $1$ T (magenta up arrows). The black down arrows indicate the position of $\nu=1$. For $p=1.0$ (Fig. \ref{fig:LLcals}(a)), $\nu=1$ occurs on the lower-field side of the crossing. At $\nu=1$, $E_F$ falls in an energy gap between two LLs and has a small but finite jump whose magnitude should give the energy gap for the $\nu=1$ integer QH state (QHS). The situation for $p=3.0$ (Fig. \ref{fig:LLcals}(c)) is qualitatively similar, except that here $\nu=1$ occurs on the higher-field side of the LL crossing. At $p=2.0$ (Fig. \ref{fig:LLcals}(b)), however, the crossing happens very close to $\nu=1$; there is no jump of $E_F$, and one would expect the absence of a $\nu=1$ QHS. The evolution presented in Fig. \ref{fig:LLcals} then predicts that the $\nu=1$ QHS should vanish and show a reentrant behavior as the density is tuned from a low to a high value. 

As illustrated in the remainder of the paper, this is very different from what we observe experimentally. Our magneto-transport measurements reveal that the LL crossing at $\nu=1$ occurs at a much higher density of $p \simeq 4.6$, as signaled by small dip in the energy gap ($^1 \Delta$) of the $\nu=1$ QHS. More important, in the entire density range, from $p \simeq 1.0$ to $\simeq 5.5$,  the measured $^1 \Delta$ is much larger than the single-particle effective Zeeman energies, and both its magnitude and field dependence ($\sim \sqrt B$) are consistent with what is expected for a QH ferromagnet with an exchange-enhanced energy gap.

The high-quality 2DHS studied here resides in a $30$-nm-wide GaAs QW grown on a GaAs (100) substrate. The QW is flanked on each side by a $510$-nm-thick Al$_{0.3}$Ga$_{0.7}$As barrier (spacer) layer followed by a C $\delta$-doping layer. The structure is then buried under a $200$-nm-thick Al$_{0.3}$Ga$_{0.7}$As layer, another C $\delta$-doping layer, and a $28$-nm-thick GaAs cap layer on top. The 2DHS has an as-grown density $p=3.8$ and a low-temperature ($T = 0.3$ K) mobility $\mu=1.3\times 10^{6}$ cm$^{2}$/Vs. We performed our measurements on a $4$ mm $\times$ $4$ mm van der Pauw geometry sample. Contacts were made by annealing InZn at $380$ \textdegree C for $70$ seconds. To tune the 2D hole density while keeping the charge distribution symmetric, we placed the sample on melted In to make a back gate and deposited Ti/Au on the top surface as the front gate. The sample was cooled down in a cryogen-free dilution refrigerator with a base temperature of $\simeq 30$ mK, and also in a $^{3}$He cryostat with a base temperature of $\simeq 0.30$ K. All transport measurements were performed using a low-frequency lock-in measurement technique at $\simeq 10$ Hz frequency with a $\simeq 50$ nA excitation current. 

\begin{figure}[t!]
  \begin{center}
    \psfig{file=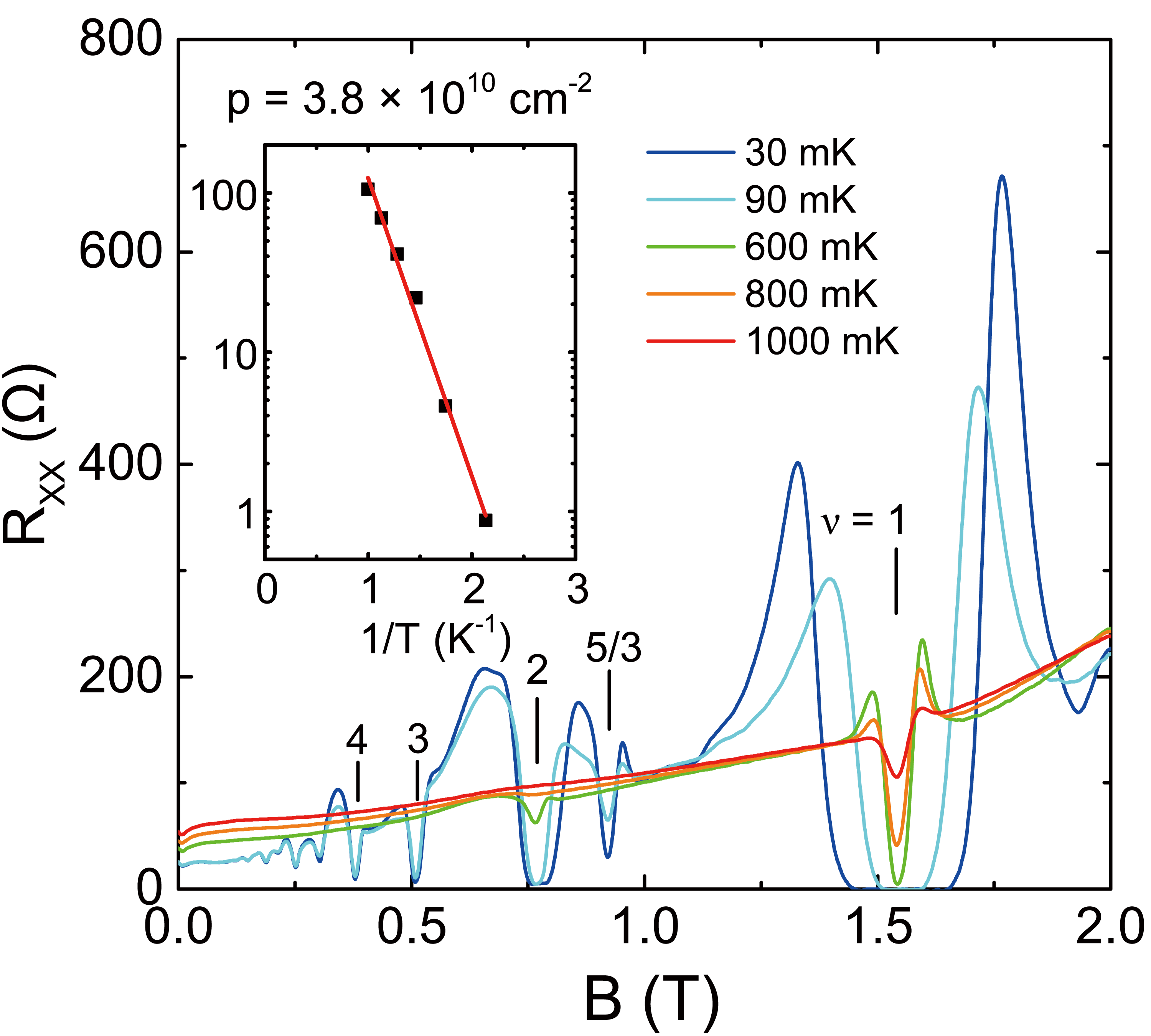, width=0.48\textwidth}
  \end{center}
  \caption{\label{transport} 
  Temperature dependence of the longitudinal resistance ($R_{xx})$ vs. $B$ at $p=3.8\times10^{10}$ cm$^{-2}$. The vertical marks indicate the field positions for $\nu=1, 2, 3, 4$, and $5/3$. The inset shows an Arrhenius plot of $R_{xx}$ minimum at $\nu=1$ vs. $1/T$, and a linear fit (in red), yielding an energy gap of $8.6$ K.
  }
  \label{fig:transport}
\end{figure}

Figure \ref{fig:transport} shows the temperature dependence of the longitudinal resistance $R_{xx}$ vs. $B$ at $p=3.8$ with both the front and back gates grounded. The vertical lines indicate the expected positions of several QHSs. The traces at the lowest temperatures show the presence of a well-developed \textit{fractional} QHS at $\nu=5/3$, as well as at 1/3, 2/5, 3/7, 2/3, 3/5, 4/7, and even at 1/5 at higher $B$ \cite{Ma.PRL.2020}; also see Supplemental Material \cite{SM}. The observation of these fractional QHSs attests to the exceptional quality of this sample, at such low density. The inset in Fig. \ref{fig:transport} shows an Arrhenius plot $R_{xx}$ at $\nu=1$ vs. $T$ from which we deduce the energy gap $^1\Delta$ for $\nu=1$. In Fig. \ref{fig:transport} inset, the red line is a linear fit through the data points, and yields $^1\Delta \simeq$ $8.6$ K.

\begin{figure}[t!]
  \begin{center}
    \psfig{file=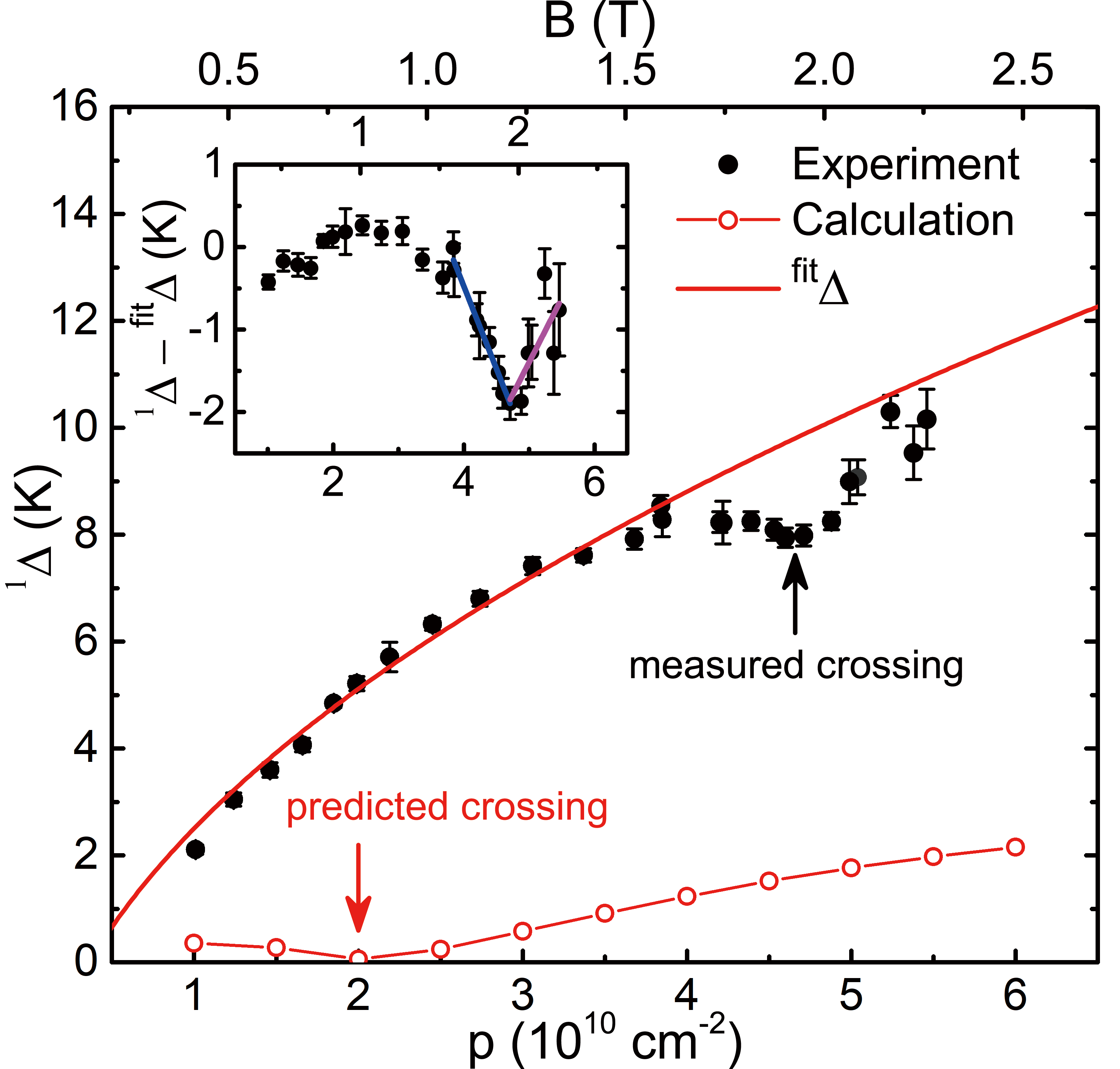, width=0.48\textwidth}
  \end{center}
  \caption{\label{summary} 
    Summary of the $\nu=1$ gap ($^1\Delta$) as a function of density (lower scale), or equivalently, $B$ (upper scale). The black circles are the experimentally measured gaps, with error bars indicated. The red open circles represent the expected gaps from LL calculations which do not include exchange energy. Data points are connected by straight lines as guides to the eye. The black arrow indicates a dip in the measured gaps, manifesting a LL crossing at $p \simeq 4.6$. The red arrow indicates the expected crossing predicted by the LL calculations. A $\sqrt{B}$ fit to the measured gaps ($^{fit}\Delta$) is shown as a red curve \cite{Footnote.sqrt.fit}. Inset: Deviation of the measured $^1\Delta$ from $^{fit}\Delta$ as a function of density (lower scale), or equivalently, $B$ (upper scale). The blue and magenta solid lines show linear fits to the data points near the measured crossing.
  }
  \label{fig:summary}
\end{figure}

We repeated the activation energy measurements at $\nu=1$ for 2D hole densities ranging from $\simeq 1.0$ to $\simeq 5.5$. This was achieved by gating the sample using both the front and the back gates while keeping the charge distribution symmetric. A summary of $^1\Delta$ as a function of $p$ is shown in Fig. \ref{fig:summary}. Error bars for the measured energy gaps, based on the Arrhenius fits to the data points (see Fig. \ref{fig:transport} inset and also \cite{SM}) are also shown. The red open circles are the gaps determined from the calculated LLs as shown in Fig. \ref{fig:LLcals}, namely the magnitude of the jump in $E_F$ at $\nu=1$. A LL crossing at $\nu=1$ is experimentally observed at a density of $p \simeq 4.6$, evinced by a dip in $^1\Delta$. A clear discrepancy is seen in Fig. \ref{fig:summary} between the predicted and measured positions of the crossing. This is not a surprise; in previous studies, similar LL calculations predict a \textit{second} LL crossing also at a somewhat lower density compared to the experiments \cite{Liu.PRB.2014, Footnote.crossing.discrepancy}. Besides the discrepancy in the predicted and measured positions of the crossing in Fig. 4, there is also a qualitative difference between the calculated and the experimentally measured gaps: The calculations predict that, starting from low densities, $^1\Delta$ should decrease as a function of increasing density, vanish when the crossing occurs at $\nu=1$, and then increase again as the density is further increased. In sharp contrast, the experimentally measured $^1\Delta$ remain finite and are significantly larger than the calculations predict at all densities. The remarkable prominence of the $\nu=1$ QHS is also evident in Fig. \ref{fig:transport} traces: At higher temperatures, e.g. at 800 mK, the only minimum seen in $R_{xx}$ is at $\nu=1$. 

We attribute the fact that the measured $^1\Delta$ is by far larger than the gap expected from the LL calculations to the presence of a QH ferromagnet at $\nu=1$ with an exchange-enhanced energy gap \cite{Footnote.nu2}. In an ideal 2D carrier system, with zero layer thickness, no LL mixing, and no disorder, the energy gap for single spin flips in a QH ferromagnet is expected to be equal to the exchange energy $E_{ex}=\sqrt{\pi/2} e^2/4 \pi \epsilon \epsilon_0 l_B$ where $\epsilon$ is the dielectric constant of the host material ($13$ for GaAs) and $l_B=\sqrt{e \hbar/B}$ is the magnetic length \cite{Sondhi.PRB.1993, Nicholas.PRB.1988, Zhang.PRB.1986}. This value is reduced by a factor of $2$ if the excitations are skyrmions \cite{Sondhi.PRB.1993}.  Note that $E_{ex}$ should vary as $\sqrt{B}$, and this is indeed what we observe in our experiments: the measured $^1\Delta$ overall follows a $\sqrt{B}$ fit except in the vicinity of the crossing where it deviates from the $\sqrt{B}$ fit (see Fig. \ref{fig:summary}). Quantitatively, for our sample, at $B=2$ T, the expected, ideal $E_{ex}$ is $\simeq 85$ K, much larger than our measured $^1\Delta \simeq 8$ K near the LL crossing. However, this is reasonable and in line with previous measurements of $^1\Delta$ at LL crossings. For example, for 2D electrons in GaAs in the limit of zero g-factor, $^1\Delta \simeq 12$ K was reported when $E_{ex} \simeq 213$ K \cite{Maude.PRL.1996}, and for AlAs 2D electrons, $^1\Delta \simeq 6$ K was quoted when $E_{ex} \simeq 263$ K \cite{Shkolnikov.PRL.2005}. The much smaller measured energy gaps compared to $E_{ex}$ have been attributed to the presence of skyrmions, finite electron layer thickness, LL mixing, and disorder, all of which are known to reduce the gap \cite{Maude.PRL.1996, Schmeller.PRL.1995, Shukla.PRB.2000, Zhao.Tongzhou.PRB.2021}. Considering these factors, the gap we measure, $^1\Delta \simeq 8$ K, is in fact surprisingly large, especially taking into account the significant amount of LL mixing we expect in our 2DHS because of its very low density and relatively large effective mass \cite{Zhu.SSC.2007, fnote1}. 

Next we discuss the role of the $\nu=1$ LL crossing on $^1\Delta$ in our 2DHS. In previous measurements of a $\nu=1$ LL crossing in GaAs and AlAs 2D electrons, a pronounced minimum in $^1\Delta$ was observed at the LL crossing \cite{Maude.PRL.1996, Shkolnikov.PRL.2005}. The minimum, and the rapid enhancement of $^1\Delta$ away from the crossing, were attributed to the predominance of skyrmions at $\nu=1$ at the crossing and the reduction in skyrmions’ size as $\nu=1$ is tuned away from the crossing. In our data, we do observe a minimum in $^1\Delta$ at the crossing but it is not as pronounced as in the above cases. It is worth emphasizing that the experiments of Refs. \cite{Maude.PRL.1996, Shkolnikov.PRL.2005} were performed at constant density and the LL crossing was induced by applying either hydrostatic pressure \cite{Maude.PRL.1996} or uniaxial strain \cite{Shkolnikov.PRL.2005}. In contrast, in our measurements we vary the density to cause the LL crossing at $\nu=1$. This implies that, if $E_{ex}$ is dominant, we would expect an increasing $^1\Delta$, reflective of the expected $E_{ex} \sim \sqrt{B}$ behavior, with a minimum, accounting for the vanishing effective Zeeman energy (and potential existence of skyrmions), superimposed on it. Our data in Fig. \ref{fig:summary} clearly exhibit the $\sqrt{B}$ dependence of the $^1\Delta$, consistent with the increase of $E_{ex}$ with $B$ \cite{Footnote.sqrt.fit}. 

In order to better present the behavior of $^1\Delta$ near the crossing, in Fig. \ref{fig:summary} inset, we show the deviation of the measured $^1\Delta$ from $^{fit}\Delta$ as a function of density (lower scale), or equivalently, $B$ (upper scale). A sharp minimum is clearly seen near the measured crossing at $p \simeq 4.6$ ($B \simeq 1.9$ T). The blue and magenta solid lines show linear fits to the data points below and above the crossing, giving slopes of $\simeq -4.8$ K/T and $\simeq 3.7$ K/T, respectively. Note that the magenta line fitting is less reliable because of the relatively larger error bars in $^1\Delta$. Ideally, the ratios of these slopes to the theoretically calculated slopes in a single-particle picture should yield a reasonable estimate of the size of skyrmions near the crossing \cite{Maude.PRL.1996, Shkolnikov.PRL.2005}. The slopes we deduce from Fig. 4 (inset) fits are indeed much larger than the slopes implied by the calculated LLs (red open circles in Fig. 4), suggesting large-size skyrmions, consistent with the conclusions reached in \cite{Lupatini.PRL.2020}. However, one should be cautious in determining a quantitative size for the skyrmions from our study, given the discrepancy between the observed and calculated positions of the LL crossing. 

We emphasize that the fact that our measured $^1\Delta$ is much larger than the effective Zeeman energy expected from the LL calculations, and its $\sqrt{B}$ dependence, are both strong indications that $^1\Delta$ is dominated by the exchange interaction and signals a robust QH ferromagnet. Our results should stimulate future many-body calculations that take the exchange energy of the 2D system into account and explain the experimental data quantitatively.

\begin{acknowledgments}

We acknowledge support by the National Science Foundation (NSF) Grants Nos. DMR 2104771 and ECCS 1906253) for measurements, the U.S. Department of Energy Basic Energy Sciences (Grant No. DE-FG02-00ER45841) for sample characterization, and the Eric and Wendy Schmidt Transformative Technology Fund and the Gordon and Betty Moore Foundation’s EPiQS Initiative (Grant No. GBMF9615 to L.N.P.) for sample fabrication. We thank J. K. Jain for illuminating discussions.

\end{acknowledgments}



\section*{Supplementary material}

# Supplemental Material to "Robust quantum Hall ferromagnetism near a gate-tuned $\nu = 1$ Landau level crossing"

Meng K. Ma,[1, *] Chengyu Wang,[1, *] Y. J. Chung,[1] L. N. Pfeiffer,[1]
K. W. West,[1] K. W. Baldwin,[1] R. Winkler,[2] and M. Shayegan[1]

[1]*Department of Electrical and Computer Engineering,
Princeton University, Princeton, New Jersey 08544, USA*
[2]*Department of Physics, Northern Illinois University, DeKalb, Illinois 60115, USA*

(Dated: October 19, 2022)

## I. DEPENDENCE OF $\nu = 1$ LANDAU LEVEL CROSSING ON QUANTUM WELL WIDTH

To further study the crossing of the lowest two Landau levels at $\nu = 1$, we calculated the density at which the Landau level crossing occurs at $\nu = 1$ as a function of the quantum well width ($w$), as shown in Fig. S1. The calculations were performed using the multiband envelope function approximation based on the $8 \times 8$ Kane Hamiltonian [2]. The black curve is the calculation for the symmetric quantum well and the red curve is for the asymmetric quantum well. The calculation for the asymmetric case assumes that all the holes in the quantum well originate from one side, i.e., that the barrier potential is flat on one side of the quantum well. As seen in the figure, the asymmetry of the potential has a small effect on the position of the crossing. In other words, the crossing is a robust phenomenon that remains present in a quantum well with asymmetric charge distribution. We emphasize that, as discussed in the main text, the experimentally observed crossing occurs at a different density compared to the prediction of the calculations. The results shown in Fig. 5 should therefore be treated cautiously and only qualitatively.

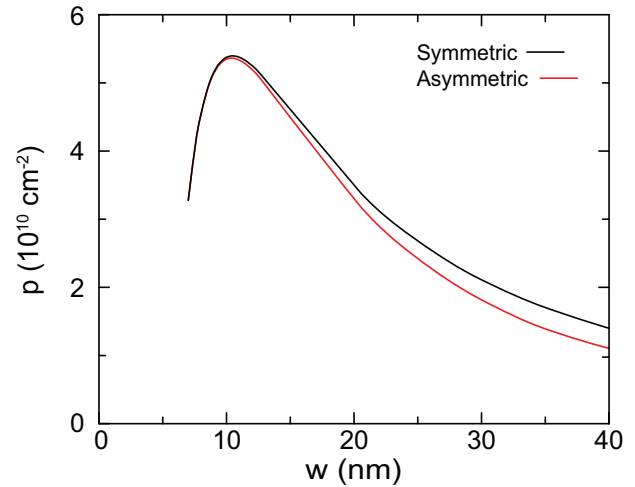

FIG. S1. Calculated density for the Landau level crossing at $\nu = 1$ as a function of the quantum well width ($w$). The black line is calculated for a symmetric quantum well and the red line is for an asymmetric quantum well, assuming single-sided doping.

## II. ADDITIONAL TRANSPORT DATA

Figure S2 shows the temperature dependence of the longitudinal resistivity ($\rho_{xx}$) vs magnetic field ($B$) at density $p = 3.8 \times 10^{10}$ cm$^{-2}$ for our two-dimensional hole system (2DHS). The measurements are performed in a cryogen-free dilution refrigerator. The $y$ scale for the grey trace is expanded by a factor of 20. The fractional quantum Hall state at $\nu = 1/3$ is fully developed at the lowest temperatures, with insulating phases, signaling the formation of pinned Wigner solid states, emerging on its flanks [1]. At high temperature (160 mK), there is even a clear minimum at $\nu = 1/5$, indicating a developing $\nu = 1/5$ fractional quantum Hall state. Other, higher order, fractional qauntum Hall states at $\nu = 2/5, 3/7, 3/5$ and $4/7$ are also seen in Fig. S2, attesting to the high quality of the sample.

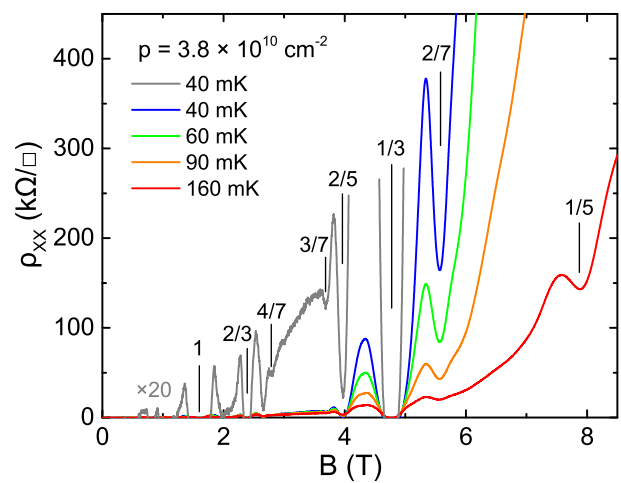

FIG. S2. Temperature dependence of the longitudinal resistivity ($\rho_{xx}$) vs magnetic field ($B$) at density $p = 3.8 \times 10^{10}$ cm$^{-2}$. The $y$ scale for the grey trace is expanded by a factor of 20.

## III. DETAILS OF $\nu = 1$ GAP MEASUREMENTS

Figure S3 shows the measured Arrhenius plots of $R_{xx}$ at $\nu = 1$ at various densities for the 2DHS studied here. The measurements are performed in a $^3$He cryostat with a base temperature of $\simeq 0.30$ K. The measured densities, derived from the field position of the integer quantum Hall states, are shown on the top-right legend in units of $10^{10}$ cm$^{-2}$. Linear fits are shown in red for each density to extract the $\nu = 1$ gaps ($^1\Delta$), which are summarized in Fig. 4 of the main text. The linear fit applies only in a certain temperature range for each density. On the higher temperature side, $R_{xx}$ starts to deviate from the linear behavior when the quantum Hall state gets weak and $R_{xx}$ tends to saturate towards a constant value. The temperature where $R_{xx}$ starts to deviate from the linear behavior progressively becomes lower with decreasing density, implying that the $\nu = 1$ quantum Hall state gets weaker monotonically with decreasing density, consistent with the measured energy gaps extracted from the fitting; see Fig. 4 of the main text. On the lower temperature side (for density $p = 3.85$ and 2.45 in Fig. S3, and $p = 3.8$ in Fig. 3 inset of the main text), the quantum Hall state almost fully develops and $R_{xx}$ becomes too small to measure accurately, and thus those data points are not included in the figure. We note that the difference between the measured energy gaps in different cool downs is very small $\lesssim 5$ %, confirming the accuracy and reproducibility of the gap measurements.

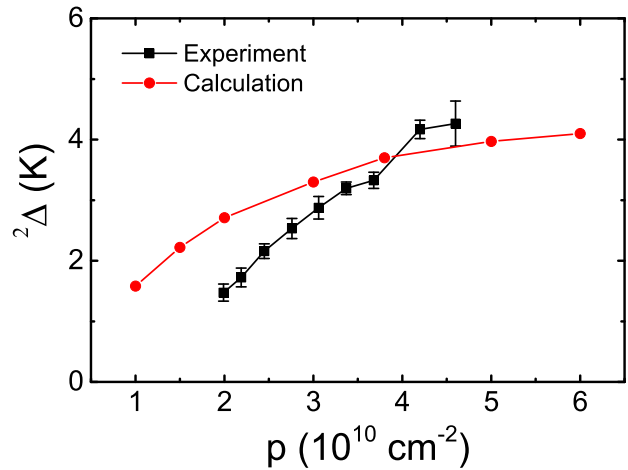

FIG. S4. Summary of the $\nu = 2$ gap ($^2\Delta$) as a function of density ($p$). The black squares are the experimentally measured values, with error bars indicated. The red circles are the expected gaps from single-particle Landau level calculations. Both sets of data points are connected by straight lines as guides to the eye.

## IV. ENERGY GAPS AT $\nu = 2$

Figure S4 shows a summary of the $\nu = 2$ gap ($^2\Delta$) as a function of the 2D hole density. The black squares are the experimentally measured gaps, obtained using the same method as used to deduce $^1\Delta$. The red circles are the gaps determined from the calculated single-particle Landau levels as shown in Fig. 2 of the main text, namely the magnitude of the jump of $E_F$ at $\nu = 2$. The points are connected by straight lines as guides to the eye. Note that the measured $^2\Delta$ are comparable to the calculated gap values in the single-particle picture, as opposed to the large enhancement of the $^1\Delta$ seen in Fig. 4 of the main text. This is because the ground state at $\nu = 2$ is *not* a quantum Hall ferromagnet when the lowest two Landau levels with different pseudo-spin species are occupied. The absence of a large enhancement of the $^2\Delta$ further attests to the significant role that exchange energy plays in $^1\Delta$.

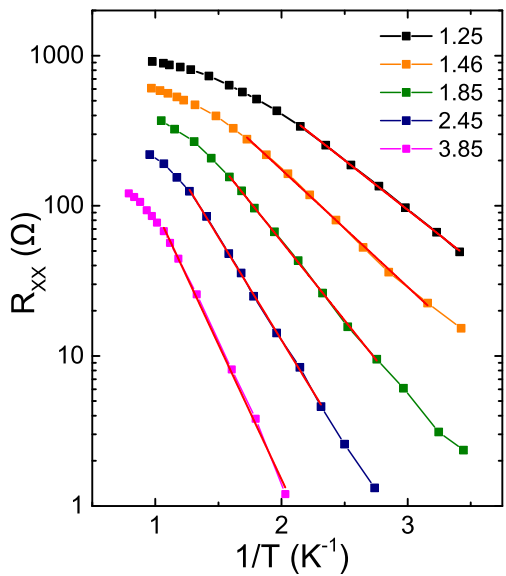

FIG. S3. Arrhenius plots of $R_{xx}$ at various densities for the 2DHS studied here. The densities labeled in the figure are in units of $10^{10}$ cm$^{-2}$. The red straight lines are linear fits, with the gap values summarized in Fig. 4 of the main text.

* These two authors contributed equally
[1] M. K. Ma, K. A. Villegas Rosales, H. Deng, Y. J. Chung, L. N. Pfeiffer, K. W. West, K. W. Baldwin, R. Winkler, and M. Shayegan, Thermal and Quantum Melting Phase Diagrams for a Magnetic-Field-Induced Wigner Solid, Phys. Rev. Lett. **125**, 036601 (2020).
[2] R. Winkler, *Spin-Orbit Coupling Effects in Two-Dimensional Electron and Hole Systems*, (Springer, Berlin, 2003).




\begin{thebibliography}{99}

\bibitem{Bloch.ZP.1929} F. Bloch, Bemerkung zur Elektronentheorie des Ferromagnetismus und der elektrischen Leitfähigkeit, Z. Phys. {\bf 57}, 545 (1929).

\bibitem{Stoner.RPP.1947} E. C. Stoner, Ferromagnetism, Rep. Prog. Phys. {\bf 11}, 43 (1947).

\bibitem{Ashcroft.Mermin.1975} N. W. Ashcroft and N. D. Mermin, \textit{Solid State Physics}, (Saunders, Philadelphia, 1976), p. 682.

\bibitem{Sharpe.Science.2019} A. L. Sharpe, E. J. Fox, A. W. Barnard, J. Finney, K. Watanabe, T. Taniguchi, M. A. Kastner, and D. Goldhaber-Gordon, Emergent ferromagnetism near three-quarters filling in twisted biayer graphene, Science {\bf 365}, 605 (2019).

\bibitem{Roch.PRL.2020} J. G. Roch, D. Miserev, G. Froehlicher, N. Leisgang, L. Sponfeldner, K. Watanabe, T. Taniguchi, J. Klinovaja, D. Loss, and R. J. Warburton, First-Order Magnetic Phase Transition of Mobile Electrons in Monolayer MoS$_2$, Phys. Rev. Lett. {\bf 124}, 187602 (2020).

\bibitem{Polshyn.Nature.2020} H. Polshyn, J. Zhu, M. A. Kumar, Y. Zhang, F. Yang, C. L. Tschirhart, M. Serlin, K. Watanabe, T. Taniguchi, A. H. MacDonald, and A. F. Young, Electrical switching of magnetic order in an orbital Chern insulator, Nature {\bf 588}, 66 (2020).

\bibitem{Hossain.PNAS.2020} Md. S. Hossain, K. A. Villegas-Rosales, M. K. Ma, Y. J. Chung, L. N. Pfeiffer, K. W. West, K. W. Baldwin, and M. Shayegan, Observation of Spontaneous Ferromagnetism in a Two-Dimensional Electron System, Proc. National Acad. Sciences (PNAS) {\bf 117}, 32244 (2020).

\bibitem{Kim&Kivelson.PNAS.2021} K. S. Kim and S. A. Kivelson, Discovery of an Insulating Ferromagnetic Phase of Electrons in Two Dimensions, PNAS {\bf 118}, e2023964118 (2021).

\bibitem{Hossain.Nat.Phys.2021}	Md. S. Hossain, T. Zhao, S. Pu, M. A. Mueed, M. K. Ma, K. A. V. Rosales, Y. J. Chung, L. N. Pfeiffer, K. W. West, K. W. Baldwin, J. K. Jain, and M. Shayegan, Bloch ferromagnetism of composite fermions, Nat. Phys. {\bf 17}, 48 (2021).

\bibitem{Hossain.PRL.2021}	Md. S. Hossain, M. K. Ma, K. A. Villegas-Rosales, Y. J. Chung, L. N. Pfeiffer, K. W. West, K. W. Baldwin, and M. Shayegan, Spontaneous Valley Polarization of Itinerant Electrons, Phys. Rev. Lett. {\bf 127}, 116601 (2021). 

\bibitem{Ando.JPSJ.1974} T. Ando and Y. Uemura, Theory of oscillatory g factor in an MOS inversion layer under strong magnetic fields, J. Phys. Soc. Jpn. {\bf 37}, 1044 (1974).

\bibitem{Giuliani.PRB.1985}	G. F. Giuliani and J. J. Quinn, Spin polarization instability in a tilted magnetic field of a two-dimensional electron gas with filled Landau levels, Phys. Rev. B {\bf 31}, 6228 (1985).

\bibitem{Sondhi.PRB.1993} S. L. Sondhi, A. Karlhede, S. A. Kivelson, and E. H. Rezayi, Skyrmions and the crossover from the integer to fractional quantum Hall effect at small Zeeman energies, Phys. Rev. B {\bf 47}, 16419 (1993). 

\bibitem{Lay.APL.1993} T. S. Lay, J. J. Heremans, Y. W. Suen, M. B. Santos, K. Hirakawa, M. Shayegan, and A. Zrenner, High-quality Two-dimensional Electron System Confined in an AlAs Quantum well, Appl. Phys. Lett. {\bf 62}, 3120 (1993).

\bibitem{Koch.PRB.1993} S. Koch, R. J. Haug, K. von Klitzing, and M. Razeghi, Suppression of the Landau-level coincidence: A phase transition in tilted magnetic fields, Phys. Rev. B {\bf 47}, 4048 (1993).

\bibitem{Maude.PRL.1996} D. K. Maude, M. Potemski, J. C. Portal, M. Henini, L. Eaves, G. Hill, and M. A. Pate, Spin Excitations of a Two-Dimensional Electron Gas in the Limit of Vanishing Land\'{e} $g$ Factor, Phys. Rev. Lett. {\bf 77}, 4604 (1996).

\bibitem{Deneshvar.PRL.1997} J. Daneshvar, C. J. B. Ford, M. Y. Simmons, A. V. Khaetskii, A. R. Hamilton, M. Pepper, and D. A. Ritchie, Magnetization Instability in a Two-Dimensional System, Phys. Rev. Lett. {\bf 79}, 4449 (1997). 

\bibitem{Girvin&MacDonald.Perspectives.1996} S. M. Girvin and A. H. MacDonald, in \textit{Perspectives on Quantum Hall Effects: Novel Quantum Liquids in Low-dimensional Semiconductor Structures}, edited by S. Das Sarma and A. Pinczuk (Wiley, New York, 1996), pp. 161-224.

\bibitem{Jungwirth.PRL.1998} T. Jungwirth, S. P. Shukla, L. Smrčka, M. Shayegan, and A. H. MacDonald, Magnetic Anisotropy in Quantum Hall Ferromagnets, Phys. Rev. Lett. {\bf 81}, 2328 (1998).

\bibitem{Piazza.Nature.1999} V. Piazza, V. Pellegrini, F. Beltram, W. Wegscheider, T. Jungwirth, and A. H. MacDonald, First-order phase transitions in a quantum Hall ferromagnet, Nature {\bf 402}, 638 (1999).

\bibitem{DePoortere.Science.2000} E. P. De Poortere, E. Tutuc, S. J. Papadakis, and M. Shayegan, Resistance spikes at transitions between quantum Hall ferromagnets, Science {\bf 290}, 1546 (2000). 

\bibitem{Girvin.Phys.Today.2000} S. M. Girvin, Spin and isospin: Exotic order in quantum Hall ferromagnets, Phys. Today {\bf 53}, 39 (2000).

\bibitem{Jungwirth.PRB.2000} T. Jungwirth and A. H. MacDonald, Pseudospin anisotropy classification of quantum Hall ferromagnets, Phys. Rev. B {\bf 63}, 035305 (2000).

\bibitem{Muraki.PRL.2001} K. Muraki, T. Saku, and Y. Hirayama, Charge Excitations in Easy-Axis and Easy-Plane Quantum Hall Ferromagnets, Phys. Rev. Lett. {\bf 87}, 196801 (2001).

\bibitem{Shkolnikov.PRL.2002} Y. P. Shkolnikov, E. P. De Poortere, E. Tutuc, and M. Shayegan, Valley Splitting of AlAs Two dimensional Electrons in a Perpendicular Magnetic Field, Phys. Rev. Lett. {\bf 89}, 226805 (2002).

\bibitem{Kellog.PRL.2004} M. Kellogg, J. P. Eisenstein, L. N. Pfeiffer, and K. W. West, Vanishing Hall Resistance at High Magnetic Field in a Double-Layer Two-Dimensional Electron System, Phys. Rev. Lett. {\bf 93}, 036801 (2004).

\bibitem{Tutuc.PRL.2004} E. Tutuc, M. Shayegan, and D. A. Huse, Counterflow Measurements in Strongly Correlated GaAs Hole Bilayers: Evidence for Electron-Hole Pairing, Phys. Rev. Lett. {\bf 93}, 036802 (2004).

\bibitem{Eisenstein.Nature.2004} J. P. Eisenstein and A. H. MacDonald, Bose–Einstein condensation of excitons in bilayer electron systems, Nature {\bf 432}, 691 (2004).

\bibitem{Shkolnikov.PRL.2005} Y. P. Shkolnikov, S. Misra, N. C. Bishop, E. P. De Poortere, and M. Shayegan, Observation of Quantum Hall Valley Skyrmions, Phys. Rev. Lett. {\bf 95}, 066809 (2005).

\bibitem{Zhang.PRL.2005} X. C. Zhang, D. R. Faulhaber, and H. W. Jiang, Multiple Phases with the Same Quantized Hall Conductance in a Two-Subband System, Phys. Rev. Lett. {\bf 95}, 216801 (2005).

\bibitem{Lai.PRL.2006} K. Lai, W. Pan, D. C. Tsui, S. Lyon, M. Mühlberger, and F. Schäffler, Intervalley Gap Anomaly of Two-Dimensional Electrons in Silicon, Phys. Rev. Lett. {\bf 96}, 076805 (2006).

\bibitem{Vakili.PRL.2006} K. Vakili, T. Gokmen, O. Gunawan, Y. P. Shkolnikov, E. P. De Poortere, and M. Shayegan, Dependence of Persistent Gaps at Landau Level Crossings on Relative Spin, Phys. Rev. Lett. {\bf 97}, 116803 (2006).

\bibitem{Zhang.PRL.2006} Y. Zhang, Z. Jiang, J. P. Small, M. S. Purewal, Y.-W. Tan, M. Fazlollahi, J. D. Chudow, J. A. Jaszczak, H. L. Stormer, and P. Kim, Landau-Level Splitting in Graphene in High Magnetic Fields, Phys. Rev. Lett. {\bf 96}, 136806 (2006). 

\bibitem{Nomura.PRL.2006} K. Nomura and A. H. MacDonald, Quantum Hall Ferromagnetism in Graphene, Phys. Rev. Lett. {\bf 96}, 256602 (2006). 

\bibitem{Padmanabhan.PRL.2010} M. Padmanabhan, T. Gokmen, and M. Shayegan, Ferromagnetic Fractional Quantum Hall States in a Valley-Degenerate Two-Dimensional Electron System, Phys. Rev. Lett. {\bf 104}, 016805 (2010).

\bibitem{Gokmen.PRB.2010} T. Gokmen and M. Shayegan, Density and Strain Dependence of $\nu = 1$ Energy Gap in AlAs Quantum wells, Phys. Rev. B {\bf 81}, 115336 (2010).

\bibitem{Young.Nat.Phys.2012} A. F. Young, C. R. Dean, L. Wang, H. Ren, P. Cadden-Zimansky, K. Watanabe, T. Taniguchi, J. Hone, K. L. Shepard, and P. Kim, Spin and valley quantum Hall ferromagnetism in graphene, Nat. Phys. {\bf 8}, 550 (2012). 

\bibitem{Parameswaran.JPCM.2019} S. A. Parameswaran and B. E. Feldman, Quantum Hall valley nematics, J. Phys.: Condens. Matter {\bf 31}, 273001 (2019).

\bibitem{Li.PRL.2020} J. Li, M. Goryca, N. P. Wilson, A. V. Stier, X. Xu, and S. A. Crooker, Spontaneous Valley Polarization of Interacting Carriers in a Monolayer Semiconductor, Phys. Rev. Lett. {\bf 125}, 147602 (2020).

\bibitem{Lupatini.PRL.2020} M. Lupatini, P. Knüppel, S. Faelt, R. Winkler, M. Shayegan, A. Imamoglu, and W. Wegscheider, Spin Reversal of a Quantum Hall Ferromagnet at a Landau Level Crossing, Phys. Rev. Lett. {\bf 125}, 067404 (2020).

\bibitem{Barrett.PRL.1995} S. E. Barrett, G. Dabbagh, L. N. Pfeiffer, K. W. West, and R. Tycko, Optically Pumped NMR Evidence for Finite-Size Skyrmions in GaAs Quantum wells near Landau Level Filling $\nu=1$, Phys. Rev. Lett. {\bf74}, 5112 (1995).

\bibitem{Winkler.Book.2003} R. Winkler, \textit{Spin-Orbit Coupling Effects in Two-Dimensional Electron and Hole Systems}, (Springer, Berlin, 2003).

\bibitem{SM} For additional relevant magneto-transport data and calculations, see Supplemental Material.

\bibitem{Footnote.color.coding} The color-coded pseudo-spins of the LLs in Figs. 1 and 2 represent the two different irreducible representations of the point group $C_2$ according to which the electron and hole eigenstates in a symmetric (or asymmetric) GaAs QW on a (001) surface transform.  States transforming according to different irreducible representations are allowed to cross as a function of $B$.

\bibitem{Graninger.PRL.2011} A. L. Graninger, D. Kamburov, M. Shayegan, L. N. Pfeiffer, K. W. West, K. W. Baldwin, and R. Winkler, Reentrant $\nu=1$ Quantum Hall State in a Two-Dimensional Hole System, Phys. Rev. Lett. {\bf 107}, 176810 (2011).

\bibitem{Liu.PRB.2015} Y. Liu, S. Hasdemir, M. Shayegan, L. N. Pfeiffer, K. W. West, and K. W. Baldwin, Unusual Landau level pinning and correlated $\nu=1$ quantum Hall effect in hole systems confined to wide GaAs quantum wells, Phys. Rev. B {\bf 92}, 195156 (2015).

\bibitem{Liu.PRB.2014} Y. Liu, S. Hasdemir, D. Kamburov, A. L. Graninger, M. Shayegan, L. N. Pfeiffer, K. W. West, K. W. Baldwin, and R. Winkler, Even-denominator fractional quantum Hall effect at a Landau level crossing, Phys. Rev. B {\bf 89}, 165313 (2014).

\bibitem{Liu.PRL.2016} Y. Liu, S. Hasdemir, L. N. Pfeiffer, K. W. West, K. W. Baldwin, and M. Shayegan, Observation of an Anisotropic Wigner Crystal, Phys. Rev. Lett. {\bf 117}, 106802 (2016).

\bibitem{Fischer.PRB.2007} F. Fischer, R. Winkler, D. Schuh, M. Bichler, and M. Grayson, Transport evidence of the lowest Landau-level spin-index anticrossing in (110) GaAs two-dimensional holes, Phys. Rev. B {\bf 75}, 073303 (2007). In their study, Fischer $et$ $al.$ report an \textit{anti-crossing} of the lowest two LLs for a GaAs 2DHS confined to a triangular well grown on GaAs (110) substrate, and near a density of $p=1.4 \times 10^{11}$ cm$^{-2}$, which is much higher than ours. In our sample, in contrast, the higher symmetry of the GaAs (001) substrate is expected to lead to a \textit{crossing} at low densities. We emphasize that the calculated level crossing in the (001) case is a robust phenomenon that persists even if the QW were not symmetric.

\bibitem{Ma.PRL.2020} M. K. Ma, K. A. Villegas Rosales, H. Deng, Y. J. Chung, L. N. Pfeiffer, K. W. West, K. W. Baldwin, R. Winkler, and M. Shayegan, Thermal and Quantum Melting Phase Diagrams for a Magnetic-Field-Induced Wigner Solid, Phys. Rev. Lett. {\bf 125}, 036601 (2020).

\bibitem{Footnote.crossing.discrepancy} As we vary density in our sample, we also observe a crossing of the lowest two LLs at $\nu=2/3$, as evinced by a disappearance of the $R_{xx}$ fractional QHS minimum. The density at which we observe this crossing ($p \simeq 2.8$) is also larger than what our LL calculations predict ($p \simeq 1.4$).

\bibitem{Footnote.sqrt.fit}  The energy vs. $B$ fit shown in Fig. 4 inset as a red curve has the form: $9.8\sqrt{B}-3.8$; the units for energy and $B$ are K and T, respectively. The negative intercept, $-3.8$ K, can be interpreted as a measure of disorder, and it’s comparable to the LL broadening expected in very high-quality GaAs 2DHSs. The $B$-dependent part of the fit can also be rewritten in terms of the Coulomb energy as $\sim 0.2 e^{2}/4 \pi \epsilon \epsilon_{0} l_{B}$. Note that the coefficient (0.2) implies that the gap is a reasonably large fraction of the Coulomb (exchange) energy, despite the relatively large LL mixing parameter ($\simeq 14$ at 2 T), and finite layer thickness of our 2DHS ($w/l_{B} \simeq 1.7$ at 2 T); $w=30$ nm is our QW width.

\bibitem{Footnote.nu2} As shown in Supplemental Material, in contrast to $^1\Delta$, the energy gaps ($^2\Delta$) measured for the $\nu=2$ QHS in our sample are rather small and comparable to those expected from the LL calculations.

\bibitem{Nicholas.PRB.1988}	R. J. Nicholas, R. J. Haug, K. v. Klitzing, and G. Weimann, Exchange enhancement of the spin splitting in a GaAs-Ga$_x$Al$_{1-x}$As heterojunction, Phys. Rev. B {\bf 37}, 1294 (1988).

\bibitem{Zhang.PRB.1986} F. C. Zhang and T. Chakraborty, Spin-1 quasiparticle and spin polarization of the ground state in the fractional quantum Hall effect, Phys. Rev. B {\bf 34}, 7076 (1986).

\bibitem{Schmeller.PRL.1995} A. Schmeller, J. P. Eisenstein, L. N. Pfeiffer, and K. W. West, Evidence for Skyrmions and Single Spin Flips in the Integer Quantized Hall Effect, Phys. Rev. Lett. {\bf 75}, 4290 (1995).

\bibitem{Shukla.PRB.2000} S. P. Shukla, M. Shayegan, S. R. Parihar, S. A. Lyon, N. R. Cooper, and A. A. Kiselev, Large skyrmions in an Al$_{0.13}$Ga$_{0.87}$As quantum well, Phys. Rev. B {\bf 61}, 4469 (2000).

\bibitem{Zhao.Tongzhou.PRB.2021} T. Zhao, Fixed-phase diffusion Monte Carlo study of activation gap and skyrmion excitations of a $\nu = 1$ system in the presence of charged impurities, Phys. Rev. B {\bf 104}, 115303 (2021).

\bibitem{Zhu.SSC.2007} H. Zhu, K. Lai, D. C. Tsui, S. P. Bayrakci, N. P. Ong, M. Manfra, L. Pfeiffer, and K. West, Density and well width dependences of the effective mass of two-dimensional holes in (100) GaAs quantum wells measured using cyclotron resonance at microwave frequencies, Solid State Commun. {\bf 141}, 510 (2007).

\bibitem{fnote1} The LL mixing parameter, quantified as the ratio of the Coulomb and cyclotron energies, is $\kappa \simeq 14$ for our sample near the LL crossing. This estimate is based on the effective mass expected in our 2DHS from cyclotron resonance measurements ($\simeq 0.5$ in unit of free electron mass, see Ref. \cite{Zhu.SSC.2007}). This is much larger than $\kappa \simeq 0.7$ for the GaAs 2D electron system of Ref. \cite{Maude.PRL.1996}. For a detailed discussion of the effective mass and LL mixing in our 2DHS, see the Supplemental Material in Ref. \cite{Ma.PRL.2020}.

\end{thebibliography}
\end{document}